\documentclass[twocolumn,aps,showpacs,pra,superscriptaddress]{revtex4-2}
\usepackage{amsmath}
\usepackage{amssymb}
\usepackage{esint}
\usepackage{graphicx}

\usepackage[normalem]{ulem}
\usepackage[colorlinks=true,allcolors=blue,bookmarks=false,pdfusetitle]{hyperref}
\usepackage{url}

\makeatletter

\usepackage{color}

\usepackage{breakurl}

\begin{document}


\title{Dynamics of thermalization of two tunnel-coupled one-dimensional quasicondensates}


\author{F. A. Bayocboc, Jr.}
\affiliation{School of Mathematics and Physics, The University of Queensland, Brisbane, Queensland 4072, Australia}
\author{M. J. Davis}
\affiliation{School of Mathematics and Physics, The University of Queensland, Brisbane, Queensland 4072, Australia}
\affiliation{ARC Centre of Excellence for Engineered Quantum Systems, School of Mathematics and Physics, The University of Queensland, Brisbane, Queensland 4072, Australia}
\author{K. V. Kheruntsyan}
\affiliation{School of Mathematics and Physics, The University of Queensland, Brisbane, Queensland 4072, Australia}

\date{\today}

\begin{abstract}
We study the non-equilibrium dynamics of two tunnel-coupled one-dimensional quasicondensates following a quench of the coupling strength from zero to a fixed finite value. More specifically, starting from two independent quasicondensates in thermal equilibrium, with initial temperature and chemical potential imbalance, we  suddenly switch on the tunnel-coupling and analyse the post-quench equilibration in terms of particle number and energy imbalances. We find that, in certain parameter regimes, the net energy can flow from the colder quasicondensate to the hotter one and is governed by the surplus of low energy particles flowing from the cold to the hot system relative to the  high-energy particles flowing in the reverse direction. In all cases, the approach to the new thermal equilibrium occurs through transient, damped oscillations. We also find that for a balanced initial state the coupled quasicondensates can relax into a final thermal equilibrium state in which they display a thermal phase coherence length that is larger than their initial phase coherence length, even though the new equilibrium temperature is higher. The increase in the phase coherence length occurs due to phase locking which manifests itself via an increased degree of correlation between the local relative phases of the quasicondensates at two arbitrary points.
\end{abstract}

\maketitle

\section{Introduction} 

The relaxation of isolated quantum many-body systems has been the subject of intense theoretical and experimental activity in recent years \cite{polkovnikov2011}. In particular, much interest has been devoted to the understanding of how quantum many-body systems evolve after a quantum quench (\emph{i.e.,} an abrupt, nonadiabatic change in a system parameter).  Generic systems have typically been found to relax to a canonically distributed thermal state, while integrable systems relax to a state described by a generalised Gibbs ensemble \cite{rigol2007,rigol2008}. 

An important physical platform that enabled fundamental advances in this research area has been that of ultracold atomic gases.  These have allowed for the experimental realization of some  paradigmatic many-body Hamiltonians, in addition to allowing for a high degree of control over system parameters and dynamics \cite{bloch2008,bloch2012}. 

The advances made in this research area are now fostering the growth of intriguing new research frontiers such as nonequilibrium quantum thermodynamics \cite{gemmer2009book,binder2018book}. In traditional thermodynamics, the simplest building block of a thermodynamic process consists of a dynamical system  coupled to a large non-dynamical bath. In quantum thermodynamics, on the other hand, one is often concerned with the dynamics of both the system and the bath, with both being treated quantum mechanically and both being not necessarily large. Examples of open questions include how quantum correlations and fluctuations of the bath, as well as the strength of the coupling between the bath and the system,  affect the system dynamics and thermodynamics. Thus, a typical setting for quantum thermodynamics can be realised by considering two coupled quantum many-body systems treated dynamically as a combined closed system.

In this paper, we study a simple example of such a setting---a pair of tunnel-coupled one-dimensional (1D) Bose gases in the quasicondensate regime, {which is characterised by suppressed density fluctuations but a fluctuating phase and hence absence of a true long-range order \cite{Petrov-2000,Esteve-2006,Jacqmin-2011}}. This system has already been studied extensively, both theoretically \cite{whitlock2003,foini2015,foini2017,vannieuwkerk2020} and experimentally \cite{hofferberth2007,betz2011,gring2012,langen2013a,langen2013b,langen2015b,schweigler2017,pigneur2018a}, in the scenario in which 
an initially single 1D quasicondensate was coherently split into two quasicondensates and allowed to relax through dephasing \footnote{{Coherent splitting here refers to the actual experimental splitting of the transverse confinement from a single- to a double-well configuration using radio-frequency adiabatic dressed potentials, whereas in theoretical studies this is usually modelled by assuming a certain initial state of the already split system.}}. {(For related studies of elongated three-dimensional Bose-Einstein condensates, see, e.g., Refs.~\cite{bidasyuk2018,momme2019,momme2020}.)} In particular, in the experiment of Hofferberth \textit{et al}. \cite{hofferberth2007}, the authors observed a fast, subexponential decay  
of relative coherence in the completely decoupled case \cite{mazets2009,stimming2011}; the dephasing between the two quasicondensate occurred on the order of milliseconds. However, in light of subsequent experiments of Ref.~\cite{gring2012}, this fast dephasing has been reinterpreted as relaxation to {a quasistationary prethermal state (that displays equilibrium-like properties \cite{Berges-2004}), rather than to the final state of a true thermal equilibrium.} 
Furthermore, in Ref.~\cite{langen2015b}, it has been shown that the steady state reached by the system due to prethermalization is described by the generalized Gibbs ensemble. {Related experimental and theoretical studies} have also shown that during the evolution to a prethermalized state, thermal correlations emerge locally and propagate through the system in a light-cone-like evolution \cite{langen2013b,geiger2014}. After reaching the prethermal state, the system then continues to dephase in a slower second light-cone-like evolution to its final thermal state \cite{langen2018}.

Here we consider the opposite dynamical scenario: rather than coherently splitting a single quasicondensate into two, we instead consider two initially independent (uncoupled) quasicondensates that are suddenly tunnel-coupled at time $t=0$ and then relax to a new common equilibrium state. More specifically, we study two parallel harmonically trapped 1D quasicondensates with an initial imbalance in their equilibrium temperature and chemical potential. The quasicondensates are then driven out of equilibrium by quenching (at time $t=0$) the tunnelling strength from zero to a finite value $J$. Using finite-temperature $c$-field techniques \cite{blakie2008}, we simulate the subsequent rephasing of the coupled system and characterize its relaxation to the final equilibrium state in terms of real-space density and momentum distributions, energy and particle number imbalances, and phase coherences. The rephasing of such a system at zero temperature has been previously studied in Ref.~\cite{Polkovnikov_2013}.

This type of a system, where two subsystems are coupled via tunneling, typically gives rise to Josephson oscillations \cite{josephson1962}. Josephson oscillations in Bose-Einstein condensates (BECs) have been studied previously both {theoretically and experimentally \cite{smerzi1997,zapata1998,raghavan1999,marino1999,giovanazzi2000,smerzi2003,grivsins2013,spagnolli2017,pigneur2018b,albiez2005,levy2007,leblanc2011}}. An interesting effect that has been predicted and observed in such tunnel-coupled systems is the phenomenon of macroscopic self-trapping \cite{raghavan1999,albiez2005}. In phase fluctuating 1D quasicondensates \cite{whitlock2003,bouchoule2005,hipolito2010,grivsins2013}, however, such an effect has been predicted to break down  \cite{hipolito2010} due to the absence of true long-range order and the enhanced role of quantum fluctuations as compared to 3D systems. Here, in the finite-temperature version of the system, we also observe the absence of macroscopic self-trapping, as expected, because of the same lack of true long-range order, albeit due to thermal fluctuations in this case.

The paper is organised as follows. In Sec.~\ref{model} we introduce the theoretical model describing two tunnel coupled 1D quasicondensates and outline the details of the  $c$-field methods used to simulate the preparation of the initial thermal equilibrium state of the system and the subsequent real-time dynamics after a sudden switching on of the tunnel coupling. In Sec.~\ref{results} we present the results of our simulations, concentrating specifically on the discussion of the dynamics of the energy and particle imbalances during relaxation of the system (Sec.~\ref{energy-particle}); particle currents in momentum space (Sec.~\ref{momentum}); characterization of the final relaxed state (Sec.~\ref{relaxed}); and the phase coherence properties of the relaxed quasicondensates (Sec.~\ref{phase}). We summarise our findings in Sec.~\ref{summary}.

\section{System and \textit{c}-field Method}
\label{model}

\subsection{The model} 

A system of two harmonically trapped 1D Bose gases coupled via tunnel-coupling can be described by the following second-quantized Hamiltonian: 
\begin{equation}\label{eq: system}
	\hat{H} = \sum_{j=1}^{2} \hat{H}_j- \hbar J \int dx \bigl[ \hat{\Psi}_{1}^{\dagger}\hat{\Psi}_{2} + \hat{\Psi}_{2}^{\dagger}\hat{\Psi}_{1} \bigr],
\end{equation}
with
\begin{equation}\nonumber
\hat{H}_j=\int dx \,\hat{\Psi}_{j}^{\dagger} \left[ -\frac{\hbar^{2}}{2m} \frac{\partial^2 }{\partial x^2} + \frac{1}{2}m\omega^{2}x^{2}  + \frac{g}{2} \hat{\Psi}_{j}^{\dagger}\hat{\Psi}_{j} \right]   \hat{\Psi}_{j}.
\end{equation}
Here, $\hat{\Psi}_{j}(x,t)$ is the annihilation operator of the $j$th ($j=1,2$) Bose gas, satisfying equal-time bosonic commutation relations $\bigl[ \hat{\Psi}_{i}(x,t),\hat{\Psi}_{j}^{\dagger}(x',t) \bigr] = \delta_{ij}\delta(x-x')$, and $m$ is the mass of a single particle.  The quantity $\omega$ is the longitudinal trap frequency which we assume is the same for both gases, $g$ is the 1D coupling strength (given by $g = 2\hbar\omega_{\perp}a$ \cite{Olshanii-1998} in the absence of confinement induced resonances, assuming the transverse confinement of both gases is also harmonic with frequency $\omega_{\perp}$, with $a$ being the 3D $s$-wave scattering length), and $J>0$ is the strength of the tunnel-coupling between the two gases.

A uniform 1D Bose gas with linear density $\rho_{j}$ at temperature $T_{j}$ can be completely characterized by two dimensionless parameters: the dimensionless interaction strength $\gamma_{j} = mg/(\hbar^{2}\rho_{j})$ and the dimensionless temperature $\mathcal{T}_{j} = 2\hbar^{2}k_{\mathrm{B}}T_{j}/(mg^{2})$ \cite{lieb1963,yang1969}. In the case of a harmonically trapped 1D Bose gas, the longitudinal trap frequency $\omega$ serves as an additional parameter needed to characterize the system. Since the density of the harmonically trapped system becomes position dependent, the interaction parameter becomes locally defined, $\gamma_{j}(x) = mg/(\hbar^{2}\rho_{j}(x))$. The  temperature $\mathcal{T}_{j}$, on the other hand, remains a global parameter for the system. For a given chemical potential $\mu_{j}$ and temperature $\mathcal{T}_{j}$, the density at the center of the trap $\rho_{0,j}=\rho_{j}(0)$ can be used to define a dimensionless interaction strength $\gamma_{0,j} = mg/(\hbar^{2}\rho_{0,j})$, which can then serve as a global interaction parameter.

In this work, we restrict ourselves to the weakly interacting regime of the coupled 1D Bose gases, which corresponds to the condition $\gamma_{0,j} \ll 1$. Moreover, we are concerned with characteristic temperatures most readily attainable in current ultracold atom experiments, at which the highly-occupied low energy modes are dominated by thermal rather than vacuum fluctuations {(see, e.g., Refs.~\cite{Amerongen2008,Jacqmin-2011,Armijo2011,bouchoule2011})}.  This corresponds to temperatures of $k_\mathrm{B}T_j\ll \sqrt{\gamma_{0,j}}\hbar^2\rho_{0,j}^2/m$, where the 1D Bose gas is in the phase-fluctuating quasicondensate regime, characterized by the thermal phase coherence length  $l^{(\phi)}_{0,j} = \hbar^{2}\rho_{0,j}/(mk_{\mathrm{B}}T_{j})$. At these temperatures, a single quasicondensate can in fact be described not by two independent dimensionless parameters, $\mathcal{T}_{j}$ and $\gamma_{0,j}$, but by a single dimensionless parameter, which is a nontrivial combination of $\mathcal{T}_{j}$ and $\gamma_{0,j}$ (see Ref.~\cite{Thomas-2021} and references therein),
\begin{equation}\label{eq: chi0}
	\chi_{0,j} = \frac{1}{2}\gamma_{0,j}^{3/2}\mathcal{T}_{j}.
\end{equation}
In terms of the original physical parameters, $\chi_{0,j} $ can be rewritten as $\chi_{0,j}~=~k_{\mathrm{B}}T_{j}/[\hbar\rho_{j,0}\sqrt{g\rho_{j,0}/m}]$.

For two tunnel-coupled quasicondensates, each having a characteristic phase coherence length of $l^{(\phi)}_{0,j}$, there is an additional length-scale in the problem associated with the coupling strength $J$. It is given by $l^{(J)} = \sqrt{\hbar/(4mJ)}$ \cite{whitlock2003} and represents the typical length-scale at which the tunnel coupling restores a spatially constant relative phase $\Delta\phi(x,t) = \phi_{1}(x,t) - \phi_{2}(x,t)$ \cite{betz2011}; in the strong cross-coupling regime, corresponding to $l^{(J)} \ll l^{(\phi)}_{0,j}$, the two quasicondensates are phase locked. This means that the distribution of $\Delta\phi$ is peaked around zero, while the local phase in each quasicondensate remains fluctuating \cite{grivsins2013}.

\subsection{The $c$-field method} 

To study the dynamics of the coupled quasicondensates we use the \textit{c}-field method, which is a proven approach to study degenerate Bose gases at finite temperatures \cite{Castin:2000,blakie2008} {(for specific applications of the $c$-field approach to 1D Bose gases, see, e.g., Refs.~ \cite{Castin:2000,Sinatra_PRL_2001,Bradley_2005,blakie2008,stimming2010,Grisins2011,stimming2011,Bradley2015,bouchoule2016,Deuar2015,Deuar2017,Deuar2018,Thomas-2021})}. In this method, the quantum field operator $\hat{\psi}_{j}(x,t)$ is decomposed into  two regions, a coherent region and an incoherent region. The coherent region contains highly occupied low-energy modes and is described by a single classical field $\psi_{j}^{(\mathcal{C})}(x,t)$ for respective ($j=1,2$) quasicondensates. On the other hand, the incoherent region contains sparsely occupied high-energy modes that act as an effective thermal bath, treated as static, with temperature $T_j$ and chemical potential $\mu_j$ governing the thermal average number of particles in the coherent region. The boundary between these two regions is defined by an energy cut-off $\varepsilon^{(\mathrm{cut})}_j$. For initially uncoupled ($J=0$) quasicondensates, each classical field $\psi_{j}^{(\mathcal{C})}(x,t)$ obeys the following stochastic projected Gross-Piteavskii equation (SPGPE) \cite{blakie2008}, for finding the initial thermal equilibrium configuration:
\begin{eqnarray}
	&~&d\psi_{j}^{(\mathcal{C})}(x,t)=\mathcal{P}^{(\mathcal{C})}\!\left\{ - \frac{i}{\hbar}\mathcal{L}^{(\mathcal{C})}_j\psi_{j}^{(\mathcal{C})}(x,t)\,dt \right. \nonumber \\
	&+& \left. \frac{\Gamma_j}{k_BT_j}(\mu_j-\mathcal{L}^{(\mathcal{C})}_j)\psi_{j}^{(\mathcal{C})}(x,t)\, dt +dW_{\Gamma_j}(x,t)\!\right\}.
	\label{SPGPE1}
\end{eqnarray}
Here, $\mathcal{P}^{(\mathcal{C})}$ is the projection operator which sets up the high-energy cutoff for the classical field region, and $\mathcal{L}^{(\mathcal{C})}_j$ is  the Gross-Pitaevskii operator
\begin{equation}\label{GP-operator}
	\mathcal{L}^{(\mathcal{C})}_j=-\frac{\hbar^2}{2m}\frac{\partial^2}{\partial x^2}+V_j(x)+g|\psi_{j}^{(\mathcal{C})}(x,t)|^2,
\end{equation}
with $V_j(x)=\frac{1}{2}m\omega_j^2x^2$ being the trapping potential (where $\omega_1=\omega_2\equiv\omega$ in this work). In addition, $\Gamma_j$ is the growth rate, whereas the last term $dW_{\Gamma_j}(x,t)$ is the associated complex white noise, with the following nonzero correlation:
\begin{equation}\label{noise-corr}
	\langle dW^*_{\Gamma_j}(x,t) dW_{\Gamma_j}(x',t)\rangle=2\Gamma_j\delta(x-x') dt.
\end{equation}

Evolution of each stochastic trajectory (initiated from a random initial noise) according to the SPGPE, for a sufficiently long time, corresponds to sampling the canonically distributed Gibbs ensemble, according to the ergodic hypothesis. Calculating expectation values of physical quantities then corresponds to evaluating ensemble averages over a large number of independent SPGPE trajectories.

In Eq.~(\ref{SPGPE1}), the numerical value of the growth rate $\Gamma_j$ is somewhat arbitrary and can be chosen for numerical convenience as it has no consequence for the final equilibrium configurations \cite{Castin:2000,bouchoule2016}. Furthermore, while the inclusion of an energy cutoff through the projection operator $\mathcal{P}^{(\mathcal{C})}$ is important in higher dimensions in order to prevent a divergence of the atomic density, its role is less crucial in 1D. This is because the classical field predictions for the atomic density do not diverge in 1D, even in the absence of energy cutoff. For this reason, our simulations were carried out without imposing the projection operator $\mathcal{P}^{(\mathcal{C})}$, in which case the cutoff is imposed merely by the finite computational basis itself. For a sufficiently large cutoff, our results did not strongly depend on the exact value of this cutoff.   

{Once the SPGPE is evolved to its final thermal equilibrium configuration, we turn on the tunnel-coupling $J$ and simulate post-quench dynamics by evolving each stochastic realization of the $c$-field in real time according to the coupled Gross-Pitaevskii equations (GPEs),}
\begin{eqnarray}
	\frac{\partial \psi_{1}^{(\mathcal{C})}(x,t)}{\partial t}&=& - \frac{i}{\hbar}\mathcal{L}^{(\mathcal{C})}_1\psi_{1}^{(\mathcal{C})}(x,t) + i J \psi_{2}^{(\mathcal{C})}(x,t), \nonumber \\
	\frac{\partial \psi_{2}^{(\mathcal{C})}(x,t)}{\partial t}&=& - \frac{i}{\hbar}\mathcal{L}^{(\mathcal{C})}_2\psi_{2}^{(\mathcal{C})}(x,t) + i J \psi_{1}^{(\mathcal{C})}(x,t).\label{GPE}
\end{eqnarray}

In Appendix \ref{dimensionless} we show how to arrive at the dimensionless form of the SPGPE, which (for a single condensate, say $j=1$) can be written in terms of the earlier introduced dimensionless parameter $\chi_{0,1}$, {Eq. \eqref{eq: chi0} (see also Ref.~\cite{Thomas-2021})}. For a harmonically trapped system, a second parameter, the dimensionless trap frequency $\tilde{\omega}$, is needed to characterize the system; this is introduced via $\tilde{\omega}=\omega t_0$, where $t_0=(\hbar^5/mg^2k_{\mathrm{B}}^2T_1^2)^{1/3}$ is the timescale defined using the temperature $T_1$ of the first quasicondensate. For two tunnel coupled quasicondensates, we also need a dimensionless coupling strength, $\tilde{J}=Jt_0$, in addition to specifying the temperature and chemical potential imbalances, which we characterize via the ratios
\begin{equation}\label{ratios}
\alpha=\frac{T_2}{T_1}\,\,\,\, \text{and}\,\,\,\, \beta=\frac{\mu_2}{\mu_1}.
\end{equation}
With these parameters, the dimensionless parameter $\chi_{0,2}$ for the second quasicondensate is related to $\chi_{0,1}$ via $\chi_{0,2} = \bigl( \alpha/\beta^{3/2} \bigr)\chi_{0,1}$, in the Thomas-Fermi limit {\cite{pethick2002book,pitaevskii2016book}, in which the mean interaction energy per particle dominates the kinetic energy, so that $\mu_j=g\rho_{0,j}$ and therefore $\chi_{0,2}/\chi_{0,1}=(T_2/T_1)(\rho_{0,1}/\rho_{0,2})^{(3/2)}=(T_2/T_1)(\mu_1/\mu_2)^{3/2}=\alpha/\beta^{3/2}$ (see also Appendix \ref{dimensionless}).}

{All numerical simulations reported below were performed with the software package XMDS2 \cite{xmds}, using the adaptive time-step Runge-Kutta (ARK45) integration algorithm.}

\section{Results and Discussion}
\label{results}

\subsection{Energy and particle exchange} 
\label{energy-particle}

\begin{figure}[tbp]
\includegraphics[width=0.85\linewidth]{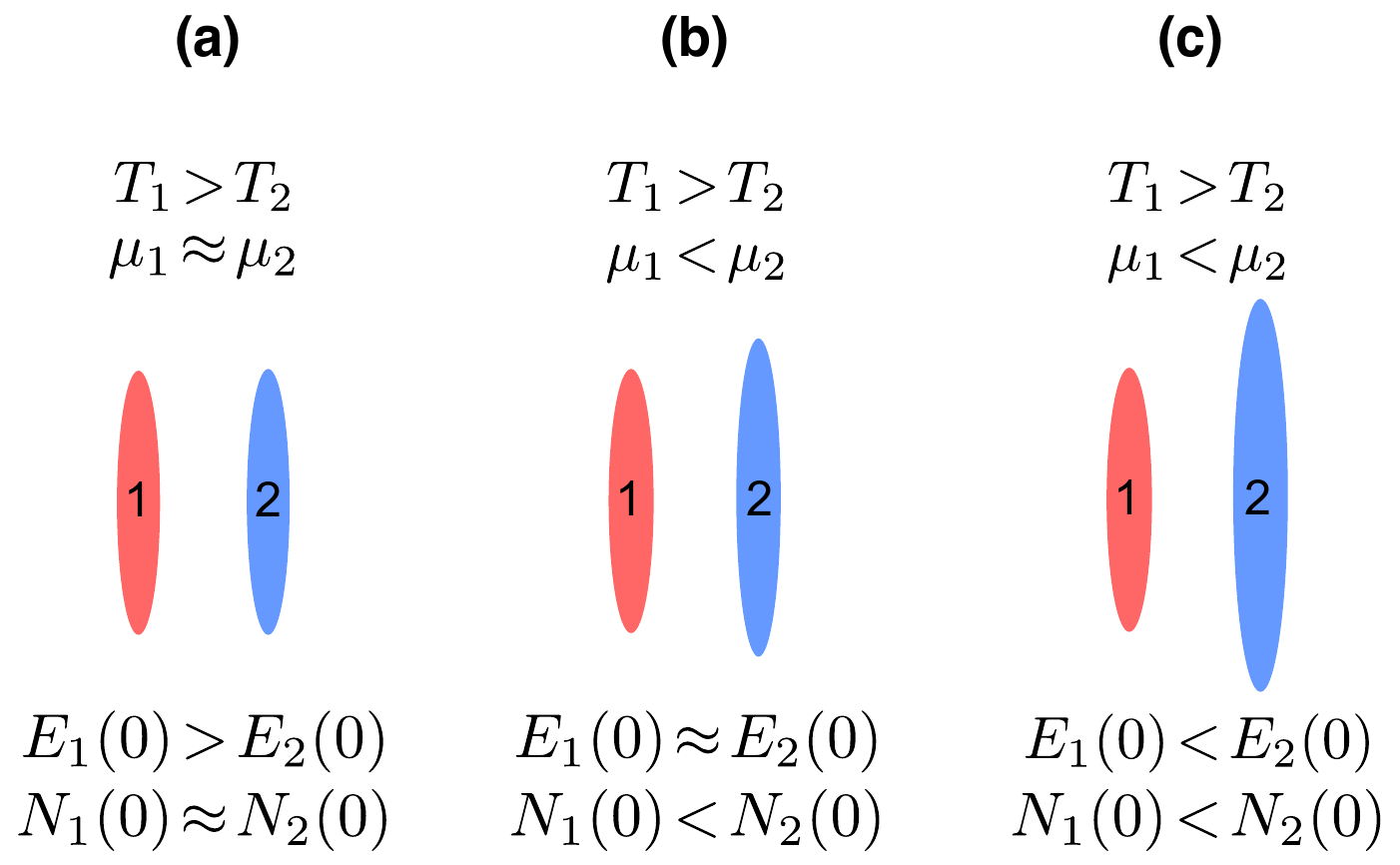}
\caption{Three different initial conditions of the quasicondensates 1 and 2, considered in the examples of Fig.~\ref{fig: DiffNT}.
\label{fig:examples}}
\end{figure}

We now consider a pair of 1D quasicondensates in parameter regimes that can be realised in ultracold atom experiments (see, \emph{e.g.}, Ref.~\cite{schweigler2017}). The initially uncoupled ($J=0$) quasicondensates are prepared independently, with generally different equilibrium temperatures $T_{j}$ ($j=1,2$) and different chemical potentials $\mu_{j}$, {at time $t=0$}. {We characterize these {initial} differences by the respective ratios, as in Eq.~\eqref{ratios}.}

At time $t=0$, the tunnel coupling $J$ is suddenly switched on to a nonzero value $J\neq0$. {This establishes thermal and diffusive contact between the two systems implying that the two systems can now exchange energy and particles while evolving to the new equilibrium state with a new common temperature and chemical potential. We characterize this exchange by the dynamics of the energy and particle number imbalances,
\begin{align} \label{E_imb}
E_{\mathrm{imb}}(t) &= \frac{E_{1}(t) - E_{2}(t)}{E_{1}(t) + E_{2}(t)},\\
N_{\mathrm{imb}}(t) &= \frac{ N_{1}(t) - N_{2}(t) }{ N_{1}(t) + N_{2}(t) }, \label{N_imb}
\end{align}
where $E_j(t)$ ($j=1,2$) are calculated as the stochastic means of the Hamiltonians for each system, in which the quantum mechanical creation and annihilation field operators are replaced by the respective complex $c$-fields, $E_j(t)=\langle \hat{H}_j\left([\psi_{j}^{(\mathcal{C})}(x,t)]^*,\psi_{j}^{(\mathcal{C})}(x,t)\right)\rangle$, and similarly for  $N_j(t)=\int dx \langle \hat{\psi}^{\dagger}_{j}(x,t) \hat{\psi}_j(x,t)\rangle = \int dx \langle |\psi_{j}^{(\mathcal{C})}(x,t)|^2\rangle $.} {We recall here that even though the coupled quasicondensates evolve according to the deterministic GPE after the quench of the tunnel coupling, the stochastic averaging here is required because each stochastic realisation (that evolves deterministically according to the GPE) starts off from a random initial condition attained after the SPGPE stage.}

\begin{figure}[tbp]
\includegraphics[width=0.92\linewidth]{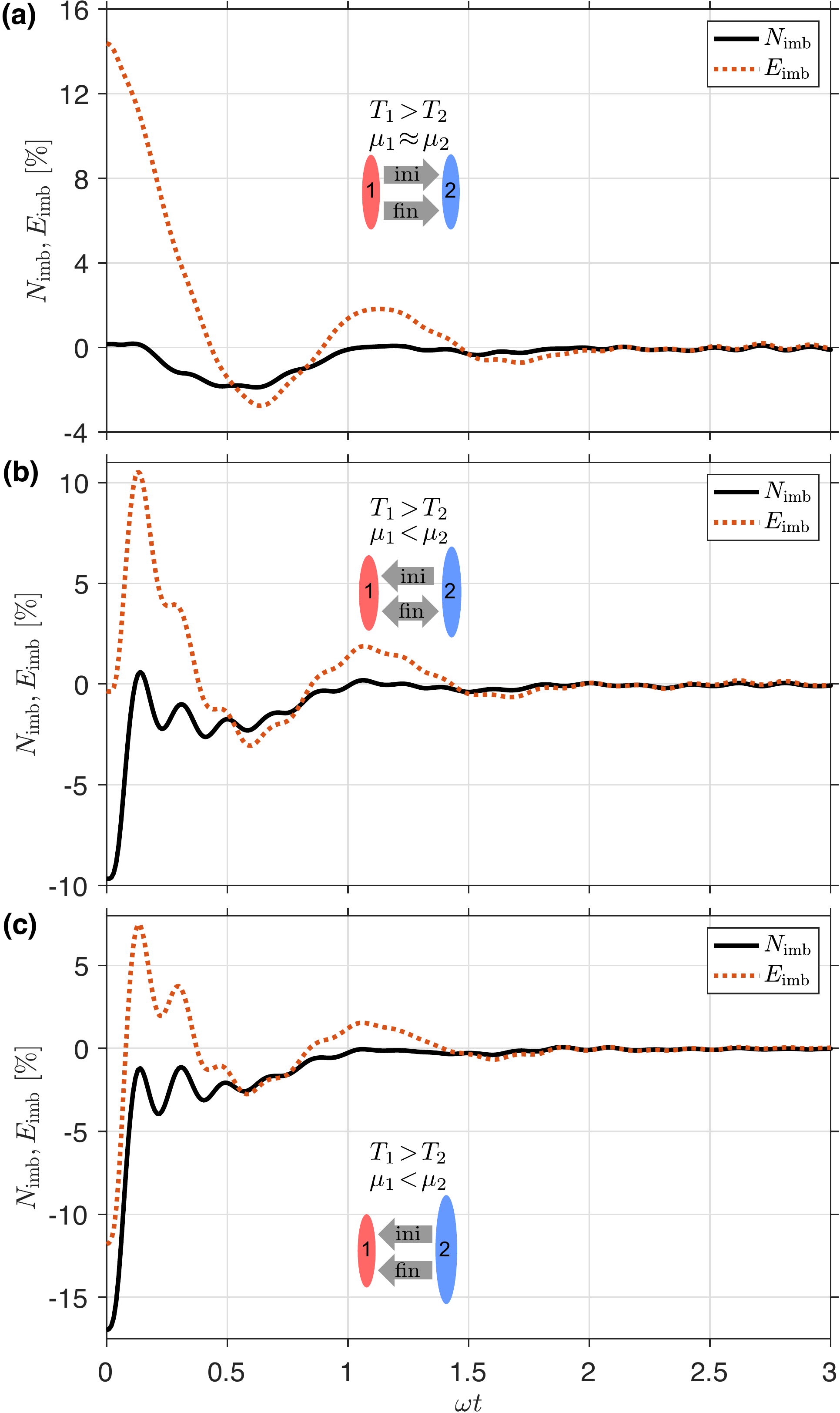}
\caption{{Time-evolution of the particle number and energy imbalances between the two quasicondensates, $N_{\mathrm{imb}}(t)$ (full, black line) and $E_{\mathrm{imb}}(t)$ (dotted, orange line), after quenching the tunnel coupling from $\tilde{J}\!=\!0$ to $\tilde{J}\!=\!3\tilde{\omega}$. The initial dimensionless parameters are $\chi_{0,1} \!\simeq \!0.138$ (or $\tilde{\mu}_1\!=\!1/\chi_{0,1}^{2/3}\!=\!3.74$)} and $\tilde{\omega}\! \simeq \!3.40\times\!10^{-2}$ in all cases; the initial ratio of temperatures is $\alpha \!=\! 0.25$; whereas the ratios of the initial chemical potentials are: (a) $\beta \!=\! 1.008$, (b) $\beta\!=\! 1.152$, and (c) $\beta \!=\!1.273$. The insets are the zoomed in parts of the curves at early times, {showing the initial direction of the energy and particle number flows, with decreasing imbalances corresponding to transfers from system 1 to system 2, and vice versa.} The red and blue ellipses illustrate the quasicondensates 1 and 2 from Fig.~\ref{fig:examples}, where we additionally show the direction of the initial (ini) and net final (fin) flow of energy by grey arrows. {In all cases, the standard errors of the means are smaller than the thickness of the lines and are not shown.}
\label{fig: DiffNT}}
\end{figure}

We consider three nontrivial scenarios, illustrated in Fig.~\ref{fig:examples}, all of which have initial temperatures $T_1>T_2$ (with the same ratio of temperatures, $\alpha=T_2/T_1=0.25$), whereas the initial chemical potentials $\mu_1$ and $\mu_2$ vary from being approximately equal to increasingly imbalanced, with $\mu_1<\mu_2$. {The evolution of the energy and particle number imbalances, $E_{\mathrm{imb}}(t)$ and $N_{\mathrm{imb}}(t)$ for these three scenarios are shown in Fig.~\ref{fig: DiffNT}.}

In the example of Fig.~\ref{fig: DiffNT}\,(a), the initial chemical potentials of the two quasicondensates are approximately the same, $\mu_1\simeq \mu_2$. 
This means that the respective initial total number of particles are also approximately the same, $N_1\simeq N_2$. {Indeed, in the quasicondensate regime that we study here the density profiles are well approximated by the Thomas-Fermi inverted parabolas, for which 
$\mu_j=g\rho_{0,j}$ and the peak densities can be expressed directly through the total atom numbers as 
$N_j$ using $\rho_{0,j}\!=\!(9m\omega^2N_j^2/32g)^{1/3}$ (see Appendix \ref{dimensionless}). 
This means that particle number imbalance is completely determined by the chemical potential imbalance, and therefore for $\mu_1\simeq \mu_2$, the initial total particle numbers will also be approximately the same so that the initial particle number imbalance is zero, $N_{\mathrm{imb}}(0)\simeq 0$.} 
{This in turn implies that after switching on the tunnel coupling $J$
the \emph{net} particle transfer from one system to the other will remain zero in the long time limit, i.e., when a new equilibrium is established and the transient oscillations in the imbalance subside back to zero. (In the examples of Fig.~\ref{fig: DiffNT}\,(a), this long time limit corresponds to approximately $t_f\sim30\omega$, which is the final time in the time-window shown.) However, the initial temperatures and energies are different, with $T_1>T_2$ and $E_1(0)> E_2(0)$ [$E_{\mathrm{imb}}(0)]>0$], implying that a net energy exchange will occur on the way to the new equilibrium.} {In this scenario, the only way the two systems can settle into a new thermal equilibrium (at a common new temperature, with zero energy imbalance in the long time limit), while maintaining the net zero particle transfer due to the already equal chemical potentials, 
is if \emph{low} energy particles predominantly tunnel from the colder system 2 to the hotter system 1, whereas the \emph{same} number of \emph{high} energy particles tunnel predominantly in the opposite direction. (See also the discussion of Fig.~\ref{fig: DensityCurrents} in the next subsection.) As we see from Fig.~\ref{fig: DiffNT}\,(a), such a net energy transfer from the initial $E_{\mathrm{imb}}(0)>0$ to final $E_{\mathrm{imb}}(t_f)\simeq0$ (where the final time $t_f$, in harmonic oscillator units, is taken to be equal to $t_f\omega=3$) takes place through transient damped oscillations. Similar transient oscillations are seen in the particle number imbalance, even though both the initial and final imbalances are zero, implying absence of any net particle transfer as expected.}

{In the example of Fig.~\ref{fig: DiffNT}\,(b), the system 1 is initially slightly hotter than the system 2 ($T_1>T_2$), but the initial chemical potentials satisfy $\mu_1<\mu_2$ and are chosen is such a way that the energies of the two systems are approximately equal, $E_1(0)\approx E_2(0)$, whereas the particle number imbalance is negative, $N_{\mathrm{imb}}(0)< 0$ ($N_1(0)<N_2(0)$). In this scenario, once the tunnel coupling is switched on, the energy initially flows from the colder system 2 to the hotter system 1, with subsequent transient damped oscillations; however, the net energy flow in the long time limit is approximately zero as both the initial and final energy imbalances are nearly zero. The net particle flow, on the other hand, is from system 2 to system 1, as expected from $\mu_1<\mu_2$. The only way this can happen is if a larger number of low energy particles flows from the colder system 2 to the hotter system 1 compared to the number of high energy particles flowing from the system 1 to system 2, leading to a net particle number flow from system 2 to system 1, while balancing out the net zero energy flow between the two systems.}

Finally, in the scenario of Fig.~\ref{fig: DiffNT}\,(c) the chemical potential imbalance ($\mu_1<\mu_2$) is {larger} than in the previous example, such that there is now an initial energy imbalance too, $E_1(0)<E_2(0)$. In this case, both the initial and long time net energy flow, as well as the particle number flow, is from the cold system 2 to the hot system 1, wherein the energy is carried by a much larger number of low energy particles flowing from system 2 to system 1, compared to the number of high energy particles flowing from system 1 to system 2.

To summarise, {in all three nontrivial examples, the low energy particles flow predominantly from the colder system 2 to the hotter system 1}, whereas the reverse is true for the high energy particles. {The net energy flow in each example depends on the overall number of low energy particles flowing from system 2 to system 1, compared to the overall number of high energy particles flowing in the opposite direction.}

\subsection{Dynamics of particles in momentum space} 
\label{momentum}

To gain further insights into the dynamics of the low and high energy particles, we now analyse the local change of the momentum distribution with time, $dn_{j}(k,t)/dt$, in each quasicondensate ($j=1,2$), where $n_j(k,t)\!=\!\iint dx \,dx'e^{ik(x-x')} \langle \hat{\psi}_{j}^{\dagger}(x,t)\hat{\psi}_{j}(x',t) \rangle$. {(Here and hereafter, we refer to $k$ as the momentum, even though it is in wave-number units, whereas the true momentum is $p\!=\!\hbar k$.)} As the dynamics of particles in each quasicondensate will generally contain contributions from the internal dynamics and tunnelling of particles from the other quasicondensate, the local rate of change $dn_{j}(k,t)/dt$ or momentum currents can be further decomposed into intra-well [$(i)$] and inter- or cross-well [$(c)$] components, 
\begin{equation}\label{eq: momentum density current}
\frac{d n_{j}(k,t)}{dt} = \frac{d n_{j}^{(i)}(k,t)}{dt} + \frac{d n_{j}^{(c)}(k,t)}{dt},
\end{equation}
with the understanding that the cross-well components have to balance each other, $dn_{1}^{(c)}(k,t)/dt = -dn_{2}^{(c)}(k,t)/dt$, due to the particle number conservation. An explicit expression for $dn_{j}(k,t)/dt$ can be calculated by using the Fourier components of the field operators, {$\hat{\psi}(x,t)=\frac{1}{2\pi}\int dk\,\hat{a}(k,t)e^{ikx}$}, and the equation of motion $i\hbar\partial\hat{\psi}_{j}(x,t)/\partial t = \bigl[ \hat{\psi}_{j}(x,t),\hat{H} \bigr]$, with the Hamiltonian $\hat{H}$ given in Eq.~\eqref{eq: system}. From this, the intra-well component is given by
\begin{equation}\label{eq: internal current}
	\frac{d}{dt}n_{j}^{(i)}(k,t) = - \frac{\partial}{\partial k} \mathcal{J}_{j}(k,t) + \frac{1}{4\pi^{2}}\iint\mathrm{d}\kappa\mathrm{d}\kappa' \mathcal{G}_{j}(k,\kappa,\kappa',t),
\end{equation}
which takes the form of a continuity equation for $n_{j}^{(i)}(k,t)$ with a nonvanishing source term (second term on the right hand side). Here,
\begin{equation}\label{eq: prob current}
	\mathcal{J}_{j}(k,t) = \frac{m\omega^{2}}{\hbar} \mathrm{Im}\biggl[ \biggl\langle \hat{a}_{j}^{\dagger}(k,t) \frac{\partial \hat{a}_{j}(k,t)}{\partial k} \biggr\rangle \biggr]
\end{equation}
is the momentum distribution flux at momentum $k$, whereas
\begin{multline}\label{eq: scattering current}
	\mathcal{G}_{j}(k,\kappa,\kappa',t) = \\
	\frac{2g}{\hbar}\mathrm{Im}\bigl[ \bigl\langle \hat{a}_{j}^{\dagger}(k,t) \hat{a}_{j}^{\dagger}(\kappa+\kappa'-k,t) \hat{a}_{j}(\kappa,t) \hat{a}_{j}(\kappa',t) \bigr\rangle \bigr]
\end{multline}
represents the scattering of two particles with initial momenta $\kappa$ and $\kappa'$, and final momenta $k$ and $\kappa+\kappa'-k$. In fluid dynamics, the second term {in the right hand side} of Eq.~\eqref{eq: internal current} represents creation of a particle with momentum $k$, hence the name -- the source term. The cross-well components, on the other hand, are given by
\begin{align}\label{eq: cross current}
\frac{d}{dt}n_{1}^{(c)}(k,t) &= 2J\,\mathrm{Im}\bigl[ \langle \hat{a}_{2}^{\dagger}(k,t) \hat{a}_{1}(k,t) \rangle \bigr], \nonumber \\
\frac{d}{dt}n_{2}^{(c)}(k,t) &= 2J\,\mathrm{Im}\bigl[ \langle \hat{a}_{1}^{\dagger}(k,t) \hat{a}_{2}(k,t) \rangle \bigr],
\end{align}
with $dn_{1}^{(c)}(k,t)/dt = -dn_{2}^{(c)}(k,t)/dt$, as expected. 

\begin{figure}[tbp]
\includegraphics[width=0.95\linewidth]{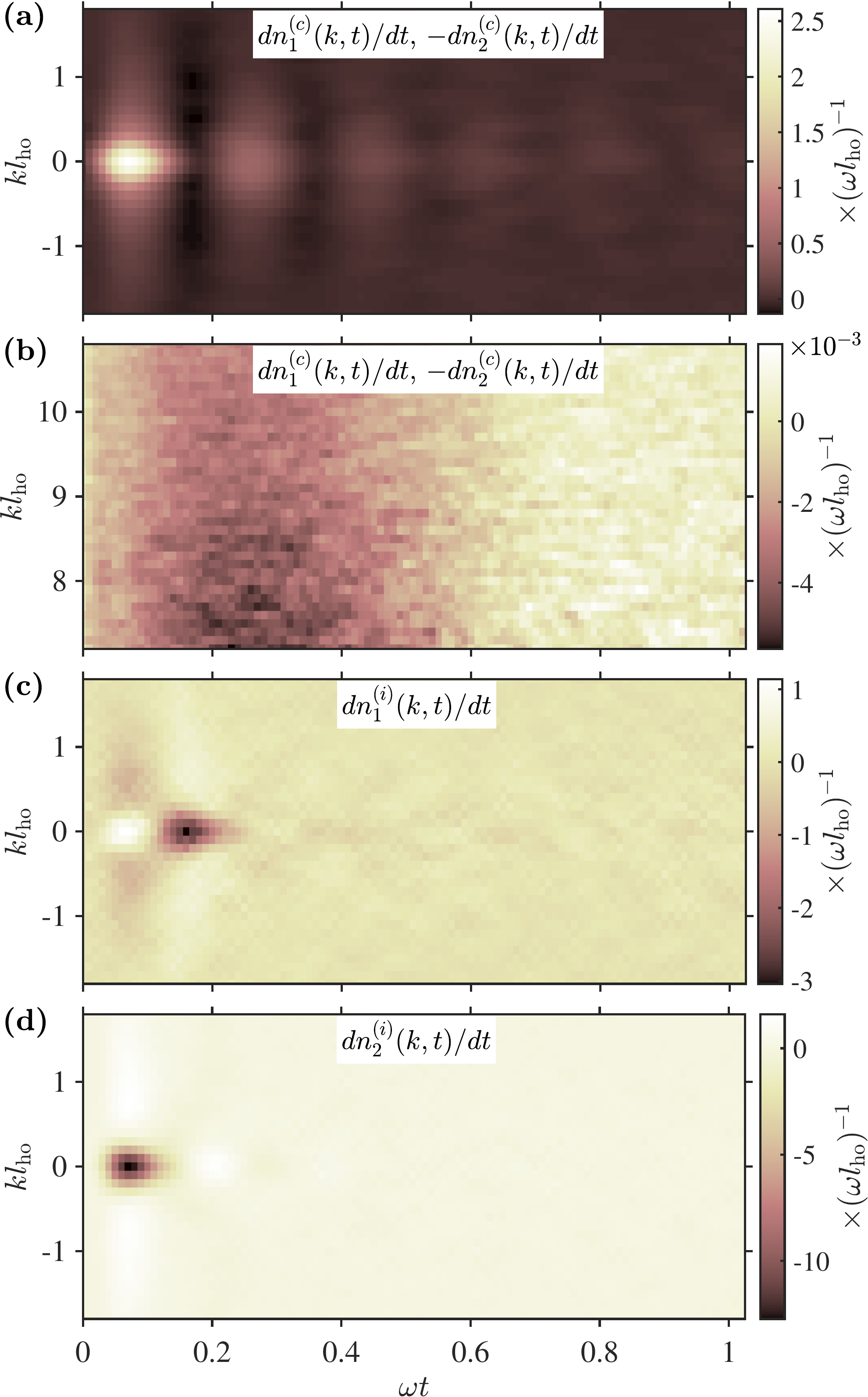}
\caption{Evolution of the cross- and intra-well components of the momentum distribution currents $dn_{j}(k,t)/dt$ versus dimensionless momentum ($k{l_{\mathrm{ho}}}$, with $l_{\mathrm{ho}}\!=\!\sqrt{\hbar/m\omega}$ being the harmonic oscillator length) and dimensionless time $\omega t$: (a) {cross-well component of the current due to tunneling of particles between the two quasicondensates, $dn_{1}^{(c)}(k,t)/dt\!=\!-dn_{2}^{(c)}(k,t)/dt$, in the low energy band; (b) cross-well component of the current in the high energy band;} (c) intra-well component $dn_{1}^{(i)}(k,t)/dt$ due to internal dynamics within quasicondensate (1); and (d) intra-well component of $dn_{2}^{(i)}(k,t)/dt$ due to internal dynamics within quasicondensate (2). Initial parameters are as in Fig.~\ref{fig: DiffNT}\,(b). {The high frequency cross-well oscillations in the low energy band occur approximately at the Josephson plasma frequency $\omega_{J}$, whereas the cross-well oscillations in the high energy band occur at a lower, Rabi frequency $\omega_R$ (see text).}
\label{fig: DensityCurrents}}
\end{figure}

{In the $c$-field approach, the  momentum currents defined above are calculated using stochastic averages over the Fourier components of $c$-field complex amplitudes in place of quantum mechanical expectation values of field operators, just like we did in the calculation of mean energies and particle numbers in Eqs.~\eqref{E_imb} and \eqref{N_imb}.}

In Fig.~\ref{fig: DensityCurrents}, we plot the cross- and intra-well components of $dn_{j}(k,t)/dt$ for two coupled quasicondensates with the same parameters as in Fig.~\ref{fig: DiffNT}\,(b). [For the parameters of Figs.~\ref{fig: DiffNT}\,(a) and \ref{fig: DiffNT}\,(c), these components have similar features as those shown in Fig.~\ref{fig: DensityCurrents} and are not plotted.] We can see here that immediately after turning on the coupling between the two quasicondensates particles start to flow across to the other quasicondensate, as well as within the same quasicondensate. Initially, there is a large flow of particles with low momenta from quasicondensate 2 to quasicondensate 1 due to the initial chemical potential difference ($\mu_{2}> \mu_{1}$). {This can be seen in Fig~\ref{fig: DensityCurrents}\,(a) as a positive valued peak in the cross-well current $dn^{(c)}_{1}(k,t)/dt$ (with $dn^{(c)}_{1}(k,t)/dt=-dn^{(c)}_{2}(k,t)/dt$) emerging approximately at $t\simeq 0.05/\omega$. The range of momenta shown here indeed corresponds to low energy particles, whereas the same cross-current dynamics in a higher energy band is shown in Fig.~\ref{fig: DensityCurrents}\,(b) where we see that the initial flow of particles if from quasicondensate 1 to quasicondensate 2, with $dn^{(c)}_{1}(k,t)/dt<0$. 
Figures ~\ref{fig: DensityCurrents}\,(c) and (d), on the other hand, show that in addition to cross-currents there is a redistribution of particles within each quasicondensate. This is seen as a nontrivial pattern of alternating regions of positive and negative intra-well currents in $dn_{1}^{(i)}(k,t)/dt$ and $dn_{2}^{(i)}(k,t)/dt$, which eventually subside to zero with time. 

The oscillations in $dn_{1}^{(c)}(k,t)/dt$ and $dn_{2}^{(c)}(k,t)/dt$) at low momenta, shown in Fig.~\ref{fig: DensityCurrents}\,(a), decay quickly---as soon as the particle numbers of the two quasicondensates become close to each other. These oscillations coincide with what we observed in the inset of Fig.~\ref{fig: DiffNT}\,(b) as high-frequency oscillations in the total particle number imbalance, which reside on top of lower frequency oscillations. The lower frequency oscillations occur at high momenta [as per Fig~\ref{fig: DensityCurrents}\,(b)] and add up to a significantly larger oscillation amplitude when integrated to give the corresponding oscillations in the total particle number imbalance of Fig.~\ref{fig: DiffNT}\,(b).} 

The way the low- and high-energy particles tunnel at different frequencies can be understood by revisiting the theory of a bosonic Josephson junction between two Bose-Einstein condensates \cite{smerzi1997,zapata1998,raghavan1999,marino1999,giovanazzi2000,smerzi2003,whitlock2003,albiez2005,bouchoule2005,levy2007,grivsins2013} and extending it to a pair of uniform 1D quasicondensates. Within the linearised theory of modulational instabilities of Josephson oscillations (where there is a small particle imbalance $|N_{\mathrm{imb}}(t)| \ll 1$ and a small relative phase $|\Delta \phi(x,t)| \equiv |\phi_{1}(x,t) - \phi_{2}(x,t)| \ll 1$), the particles oscillate between the quasicondensates with frequency given by $\omega_{\mathrm{lin}} = \sqrt{(2g\rho_{0,\mathrm{tot}}J/\hbar) + 4J^{2}}$ \cite{smerzi1997}, where $\rho_{0,\mathrm{tot}} = \rho_{0,1}+\rho_{0,2}$. Here, we can see that if the {intra-well interaction energy} is sufficiently large such that $g\rho_{0,\mathrm{tot}} \gg \hbar J$ the particles will oscillate between the two quasicondensates at a rate given by the Josephson plasma frequency $\omega_{J} = \sqrt{2Jg\rho_{0,\mathrm{tot}}/\hbar}$ \cite{Pitaevskii2001,whitlock2003}. As this condition in our inhomogeneous quasicondensate is satisfied in the high-density bulk of the system, which is also where the low-energy particles are localised, we do indeed observe that the low-$k$ components shown in Fig.~\ref{fig: DensityCurrents}\,(a) oscillate approximately at the said plasma frequency $\omega_{J}$, which is the high frequency component.  In the opposite limit where interaction between particles is negligible, i.e. for high-energy tails of the momentum distribution in our system, the particles undergo a Rabi-like oscillations with frequency given by $\omega_{R} = 2J$ \cite{smerzi1997,raghavan1999}, which is the low frequency component {shown in Fig.~\ref{fig: DensityCurrents}\,(b)}.

\subsection{Equilibrium chemical potential and temperature after relaxation} 
\label{relaxed}

We now characterize the final relaxed state of the two tunnel-coupled 1D quasicondensates. More specifically, we let the system evolve until it relaxes at some final time $t=t_{f}$ and then compute the final chemical potential and final temperature of both quasicondensates by comparing the momentum distribution of the relaxed state with that of a thermal Gibbs distribution. Given that the initial state of the system (before the coupling $J$ is turned on) is an excited state with respect to the \emph{coupled} system at $J\neq 0$, we expect the two coupled quasicondensates to relax to a final equilibrium state at a higher temperature.

Given the peak density $\rho_{j}(x=0,t_{f})\equiv \rho_{0,j}^{(f)}$ of the final relaxed state, the global equilibrium chemical potential of each quasicondensate can be estimated as $\mu_{j}^{(f)} = g \rho_{0,j}^{(f)}$ in the trap centre, using the Thomas-Fermi approximation. The temperature of the final relaxed state $T_{1}^{(f)} = T_2^{(f)}$, on the other hand, is obtained by fitting the momentum distributions $n_{j}(k,t_{f})$ of the relaxed state to an equilibrium thermal momentum distributions $n_{j}^{(\mathrm{th})}(k)$. {The thermal momentum distributions $n_{j}^{(\mathrm{th})}(k)$ are obtained by performing SPGPE simulations of two initially coupled quasicondensates according to Eqs.~\eqref{SPGPE1} with an added tunnel coupling term as in the GPE \eqref{GPE}} (i.e., by modelling the respective thermal equilibrium state after cooling down two pre-coupled quasicondensates in a transverse double well) with chemical potential $\mu_{1}^{(f)}=\mu_{2}^{(f)}=\mu$ and scanning for different values of temperature $T_{1}^{(f)} = T_2^{(f)}= T_{\mathrm{th}}$ until the obtained {equilibrium SPGPE momentum distributions match the momentum distributions of our final (time-evolved) relaxed state.}

To quantify the proximity of $n_{1}^{(\mathrm{th})}(k)$ to the final relaxed momentum distribution, $n_{1}(k,t_{f})=n_2(k,t_{f})$, we calculate the Bhattacharyya statistical distance \cite{bhattacharyya1943}
\begin{equation}\label{eq: Bhattacharyya distance}
D_{B} = -\ln[B(P,P')],
\end{equation}
where $B(P,P')$ is the Bhattacharyya coefficient given by
\begin{equation}
{B(P,P') = \sum_{\{y\}} \sqrt{P(y)P'(y)}.}
\end{equation}
Here, {$P(y)$ and $P'(y)$ are two normalized probability distributions functions of a discrete variable $y$ within the same domain}, which are being compared. As $P(y)$ approaches $P'(y)$, the Bhattacharyya coefficient tends to unity, $B(P,P')~\to~1$, which gives a distance of $D_{B} = 0$ indicating a complete overlap of the two distributions. In our case, the roles of the two distribution functions are taken by 
the normalised momentum distributions of the relaxed state, {$P_{1}(k,t_{f}) = n_{1}(k,t_{f})/\sum_{k=-k_{\max}}^{k_{\max}}n_{1}(k,t_{f})$}, and similarly for the thermal equilibrium state, {$P_{1}^{(\mathrm{th})}(k) = n_{1}^{(\mathrm{th})}(k)/\sum_{k=-k_{\max}}^{k_{\max}}n_{1}^{(\mathrm{th})}(k)$}, where the summations are over the discrete computational lattice points in momentum space.

\begin{figure}[tbp]
\includegraphics[width=0.95\linewidth]{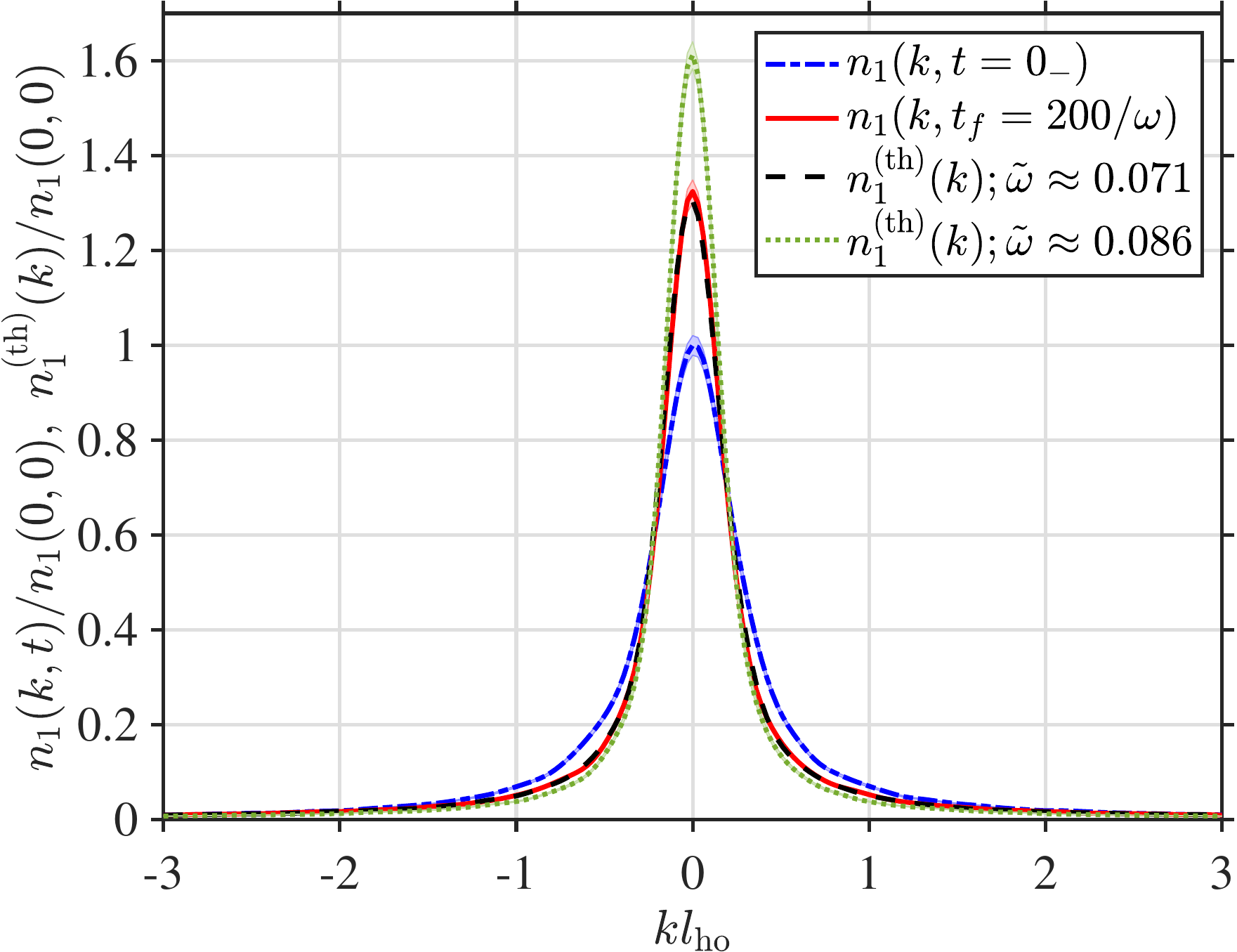}
\caption{{Momentum distributions of two initially independent but {fully} balanced ($\alpha\!=\!1$, $\beta\!=\!1$) 1D quasicondensate before and after relaxation, for $\tilde{J}=3\tilde{\omega}$. We show only the momentum distribution of the first quasicondensate $n_1(k,t)$, with the understanding that $n_2(k,t)=n_1(k,t)$ in this fully balanced case. The dimensionless momentum $kl_{\mathrm{ho}}$ for the horizontal axis is defined as in Fig.~\ref{fig: DensityCurrents}. The initial parameters are $\tilde{\omega} \!=\! 0.0857$ and $\chi_{0,1}\! =\!0.0341$,  with the respective dimensionless chemical potentials given by $\tilde{\mu}_1\!=\!1/\chi_{0,1}^{2/3}\!=\!9.50$, according to Eq.~\eqref{eqn:mu_bar} of Appendix \ref{dimensionless}, and $\tilde{\mu}_2\!=\!\tilde{\mu}_1$ (for $\beta\!=\!1$).
The two initially uncoupled quasicondensates [blue, dashed-dotted line, with $n_1(k,t\!=\!0_{-})\!=\!n_2(k,t\!=\!0_{-})$] evolve in a coupled setup and eventually relax into a new equilibrium state, with the respective momentum distribution $n_1(k, t_f\!=\!200\omega)\!=\!n_2(k,t_f=200/\omega)$ (i.e., after 200 oscillation cycles) shown as red full line.  The dashed black line is an actual thermal (Gibbs) equilibrium distribution of pre-coupled 1D quasicondensates fitted to the time-evolved, relaxed distribution; the best-fit temperature for this thermal distribution corresponds to $\tilde{\omega}\!=\! 0.0712$ (recall that $\tilde{\omega} \!\propto\! T_1^{-2/3}$), which means that the temperature of the relaxed state is higher than the initial temperature $T_1$ by a factor of $1.32$ [corresponding to $(0.0857/0.0712)^{3/2}$]. 
The new (relaxed) value of the parameter $\chi_{0,1}$ is also a factor of $1.32$ larger than initially, according to $\chi_{0,1}\!=\!4\sqrt{2}k_BT_1/(3N_1\hbar\omega)$.} The threshold for Bhattacharyya distance used here to terminate the search for the best-fit temperature between the relaxed and thermal distributions is $D_{B}\!=\!0.0001$. We also show for comparison (green dotted line) the momentum distribution of a thermal equilibrium state of an initially coupled system, but at the same initial temperature $\tilde{\omega} \!=\! 0.0857$ as the pre-quenched (lower temperature) uncoupled system; the distribution is narrower than that of the final relaxed state of the quenched system, as expected. {The light shaded areas on all curves indicate the standard errors of the means.} }
\label{fig: momdens}
\end{figure}

In Fig.~\ref{fig: momdens}, we plot the momentum distribution $n_1(k,t)$ of the first quasicondensate of a system with two initially uncoupled but {fully balanced quasicondensates, i.e., with $T_1\!=\!T_2$ ($\alpha\!=\!1$) and $\mu_1\!=\!\mu_2$ ($\beta\!=\!1$), and therefore} $N_{\mathrm{imb}}(0)\! =\! 0$ and $E_{\mathrm{imb}}(0) \!=\! 0$. [The momentum distribution for the second quasicondensate is essentially the same, {$n_{2}(k,t)\!=\!n_{1}(k,t)$}, in this balanced case.] The curves plotted are the momentum distribution before coupling, $n_{1}(k,t\!=\!0_{-})$, and at time $t_f\!=\!200/\omega$, $n_{1}(k,t_f\!=\!200/\omega)$, when the coupled system has already relaxed. {The relaxed distribution (red, full line) is fitted with the momentum distribution at thermal equilibrium of an equivalent pre-coupled system with the same number of particles  (black, dashed line), but at a higher temperature than the initial temperature of our system. As we see, the relaxed and best-fit thermal distributions are in excellent agreement with each other} (the corresponding Bhattacharyya distance here is $D_B=0.0001$), implying that the initially uncoupled quasicondensates have relaxed to a higher temperature thermal equilibrium state; {the best-fit equilibrium temperature for the relaxed distribution corresponds to $\tilde{\omega}= 0.0712$ (where we recall that $\tilde{\omega} \propto T^{-2/3}$), whereas the initial value of this parameter was $\tilde{\omega} = 0.0857$.}  We thus conclude that the final relaxed state of the coupled system is described by the thermal Gibbs ensemble, characterized only by the global temperature and the global chemical potential. In addition, we also plot the momentum distribution at thermal equilibrium of a pre-coupled system at the same temperature as the initially uncoupled system. As expected, this distribution is narrower than that of the final relaxed states as it corresponds to a lower temperature thermal equilibrium state.

\subsection{Equilibration and thermal phase coherence} 
\label{phase}

In this subsection we characterize the coherence properties of the coupled quasicondensates before and after relaxation. 
As we saw in Fig.~\ref{fig: momdens}, the final relaxed state of each quasicondensate is characterized by the momentum distribution that is narrower and has a higher population of the low-momentum modes around $k=0$ than the respective initial momentum distribution before the coupling $J$ is switched on. This observation suggests that the condensate fraction (as per Penrose-Onsager criterion \cite{Penrose-Onsager}), and hence the phase coherence of both quasicondensates, increase despite the fact that the quasicondensates have relaxed to a thermal equilibrium state at a higher, rather than lower, temperature. To confirm this, we compute 
the normalised first-order correlation function for each quasicondensate
\begin{equation}\label{eq:g1}
g_{j}^{(1)}(x,x',t) \equiv \frac{\langle \hat{\psi}_{j}^{\dagger}(x,t) \hat{\psi}_{j}(x',t) \rangle}{\sqrt{\rho_{j}(x,t)\rho_{j}(x',t)}}.
\end{equation}
and check whether the phase coherence length indeed increases after relaxation.

As a guide to evaluating the phase coherence length, we note that within {the Luttinger liquid approach for a uniform system \cite{haldane1981,Mora-Castin-2003,cazalilla2004,Bouchoule-2012} (which treats only the long wavelength excitations, for relative distances much larger than the healing length $l_j^{(h)}=\hbar/\sqrt{mg\rho_j}$, and ignores density fluctuations), the equilibrium first-order correlation function in the quasicondensate regime is expected to decay exponentially as a function of the relative distance (see, \emph{e.g.}, \cite{Bouchoule-2012} for further details),
\begin{equation}\label{eq: g1_exp}
g_{j}^{(1)}(x,x')  = e^{-|x-x'|/2l_{j}^{(\phi)}},\,\,\,\,|x-x'|\gg l_j^{(h)},
\end{equation}
where $l_{j}^{(\phi)} = \hbar^{2}\rho_{j}/mk_{\mathrm{B}}T_{j}$ is the thermal phase coherence length of the $j$th quasicondensate at density $\rho_j$ and temperature $T_{j}$.}

\begin{figure}[tbp]
\includegraphics[width=0.95\linewidth]{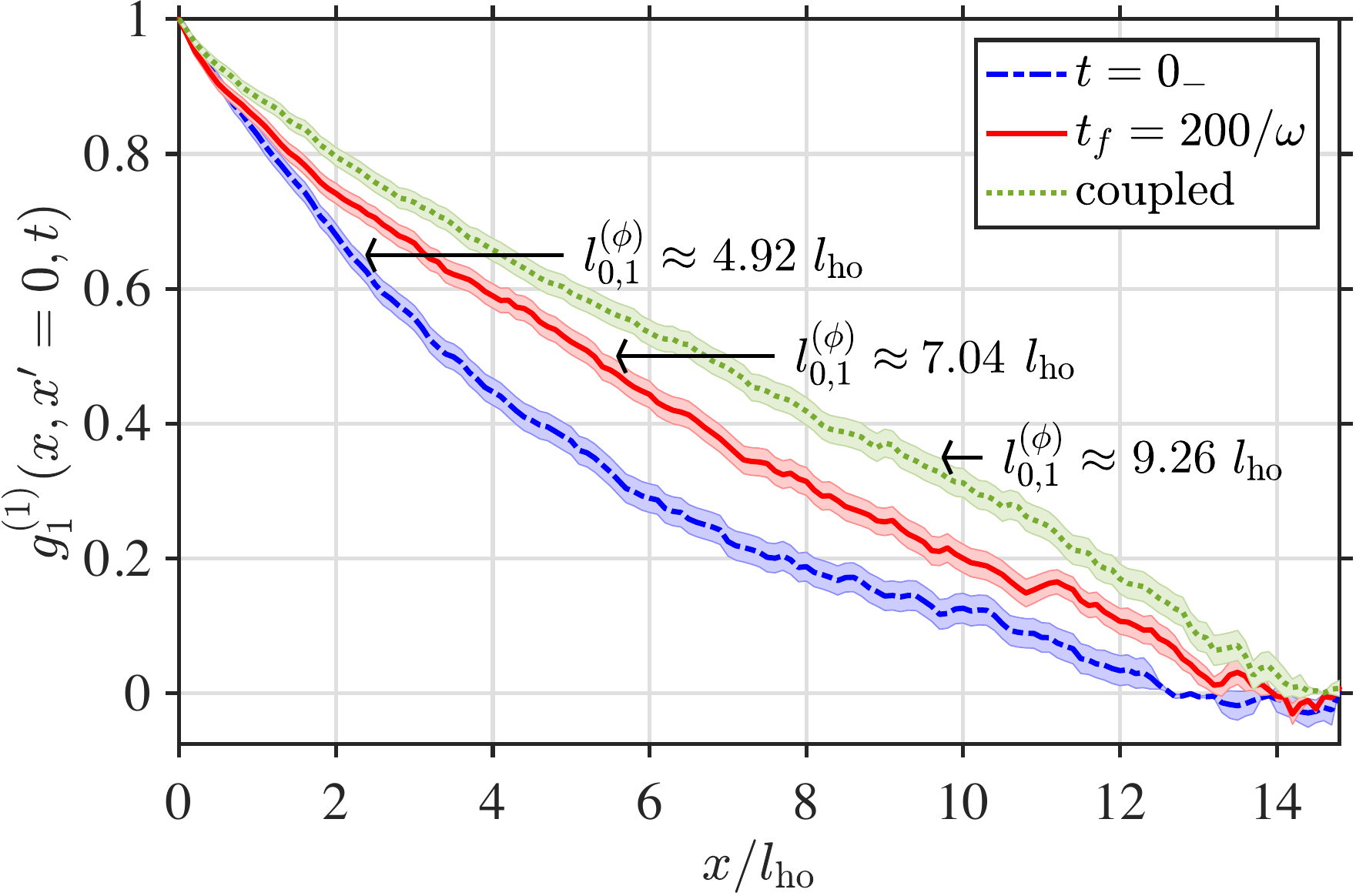}
	\caption{First-order correlation function $g_{1}^{(1)}(x,x'\!=\!0,t)$, {Eq.~\eqref{eq:g1}}, before ($t\!=\!0_{-}$, $\tilde{J}\!=\!0$, blue dash-dotted line) and after ($t\!=\!t_f$, $\tilde{J}\!=\!3\tilde{\omega}$, red full line) relaxation, for two initially independent {but fully balanced quasicondensates}. All parameters are the same as in Fig.~\ref{fig: momdens}. We also show (by the green dotted line) the thermal equilibrium correlation function $g_{1}^{(1)}(x,x'=0)$ for an initially coupled system if it were prepared (cooled down in a transverse double well) at the same temperature as the two initially independent 1D quasicondensates.  {The light shaded areas on all curves indicate the standard errors of the means.} 
	\label{fig: g1}}
\end{figure}

Adopting this result for our nonuniform system within the local density approximation and evaluating $g_{j}^{(1)}(x,x',t)$ in the trap centre, i.e., at $x'=0$ and as a function of $x$, we expect to observe exponentially decaying correlation functions, with the phase coherence length $l_{0,j}^{(\phi)}$ evaluated at the peak density $\rho_{0,j}$ and the respective temperature of the system $T_{j}$. More specifically, we calculate $g_{j}^{(1)}(x,x'=0,t)$ at time $t=0_{-}$ and at $t=t_f$ once the system has relaxed. The results of SPGPE simulations for the {initial correlation $g_{j}^{(1)}(x,x'=0,t=0_{-})$, and SPGPE plus GPE simulations for the final $g_{j}^{(1)}(x,x'=0,t=t_f)$ correlation,} are shown in Fig.~\ref{fig: g1} for the same initially {fully balanced quasicondensates and same parameters as in Fig.~\ref{fig: momdens}}. {[The first-order correlation function for the second quasicondensate, $g_{2}^{(1)}(x,x'=0,t)$, is essentially the same in this balanced case, within the statistical noise, and is not plotted for clarity.]} We observe a notable increase of the thermal phase coherence length of the quasicondensates after relaxation (red, full line) compared to that of the initial lower temperature state (blue, dash-dotted line). {The numerical values for $l_{0,1}^{(\phi)}=l_{0,2}^{(\phi)}$ shown on the figure are extracted from exponential fits, Eq.~\eqref{eq: g1_exp}, to the respective curves within the range $x/l_{\mathrm{ho}}\in [0,10]$ covering the bulk of the quasicondensate.} For comparison, we also plot (green, dotted line) the thermal equilibrium first-order correlation function of initially pre-coupled quasicondensates at the same temperature as the initially uncoupled system, {calculated in the same manner as the thermal momentum distribution $n_{j}^{(\mathrm{th})}(k)$ in Fig.~\ref{fig: momdens}, i.e., using tunnel coupled SPGPEs}. The coherence length of such a system is larger than that of the relaxed state of the post-quench coupled system, reflecting the fact that the post-quench state, immediately after turning on the tunnel coupling $J$, is an excited state of the transverse double well potential, which relaxes to a new higher temperature equilibrium state of that potential. 

The increase in the  phase coherence length after relaxation of the coupled system to a new thermal equilibrium state at a higher temperature can be explained by the fact that the tunnel coupling between the condensates is equivalent to an introduction of an effective additional degree of freedom, which generally favours establishment of long range order. This in turn is similar to the general trend that mean-field theories perform better when going from lower to higher dimensional systems or when increasing the coordination number in a many-body system.

\begin{figure}[tbp]
\includegraphics[width=0.95\linewidth]{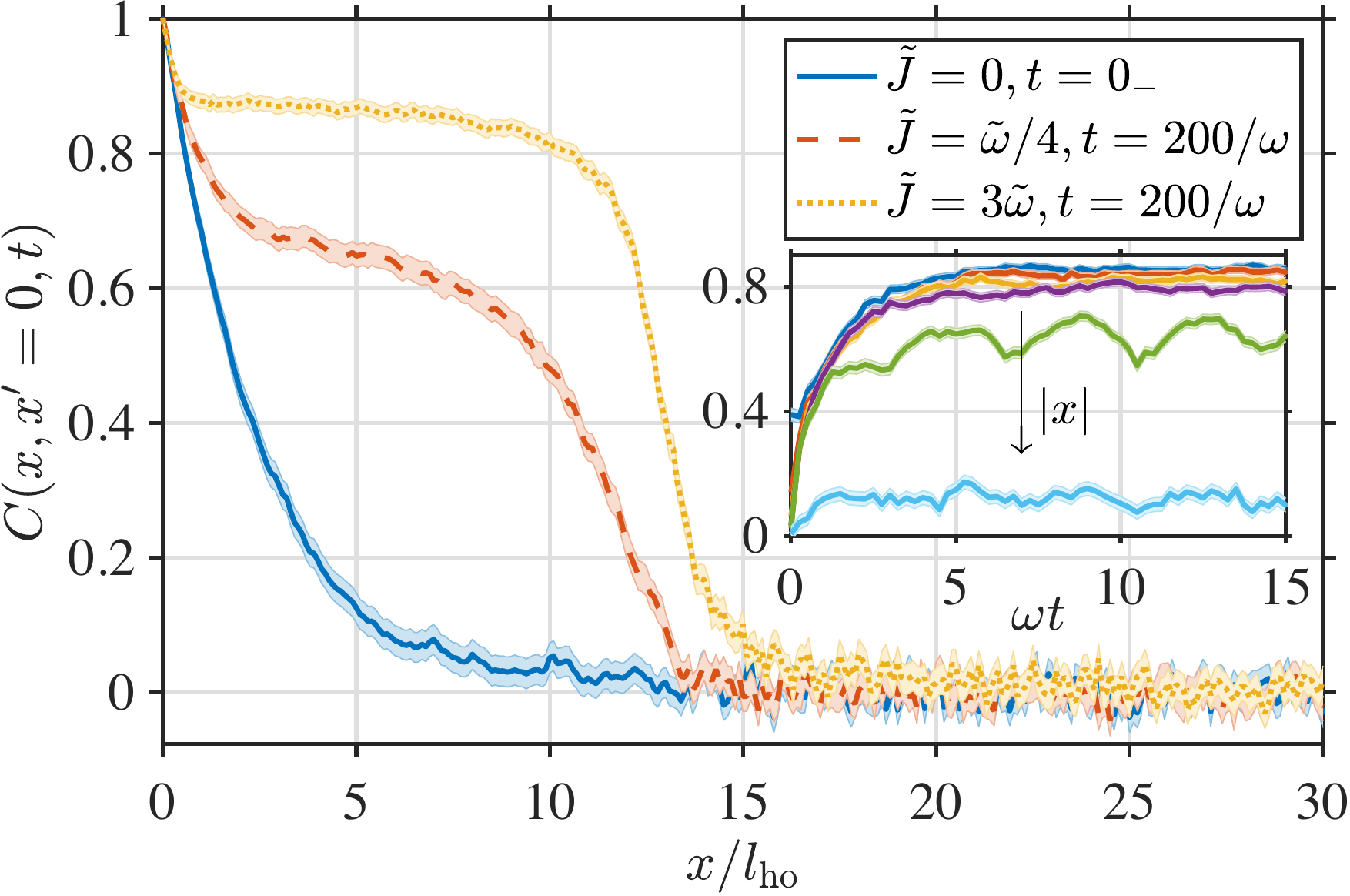}
\caption{Two-point correlation function $C(x,x'=0,t)$ as a function of distance $x$ from the trap center ($x'=0$), for two initially independent {but fully balanced quasicondensates, and the same parameters as in Fig.~\ref{fig: momdens}. The different curves correspond to different time instances: before the switching on the tunnel coupling (blue solid line, $t=0_-$) and after the coupled system has relaxed (dashed red and dotted yellow lines, $t=200/\omega$) for two different coupling strengths as indicated by the labels. The light shaded areas indicate the standard errors of the means.} 
The inset shows the evolution of $C(x,x'=0,t)$ with dimensionless time $\omega t$ (for $\tilde{J}=3\tilde{\omega}$) at several fixed points $x$, starting from $x = 2.4l_{\mathrm{ho}}$ (topmost curve) to $x = 14.4l_{\mathrm{ho}}$ (bottommost curve) with an increment of $\Delta x = 2.4l_{\mathrm{ho}}$. \label{fig: DiffNT-C2}}
\end{figure}

From the microscopic dynamical viewpoint, the increase in the phase coherence within each relaxed quasicondensate can be attributed to the phase correlation established between the quasicondensates by the tunnel coupling. A quantitative measure of such a cross-correlation, namely between the phases of the two quasicondensates, is given by the following two-point second-order correlation function  \cite{langen2015b,foini2015}:
\begin{equation}\label{eq: C2}
C(x,x',t) = \frac{\langle \hat{\psi}_{1}(x,t)\hat{\psi}_{2}^{\dagger}(x,t)\hat{\psi}_{1}^{\dagger}(x',t)\hat{\psi}_{2}(x',t) \rangle}{\rho_{1}(x,t) \rho_{2}(x',t)}.
\end{equation}
The cross-correlation $C(x,x',t)$ characterizes the degree of correlation between the complex-valued Bose fields $\hat{\psi}_{1}(x,t)$ and $\hat{\psi}_{2}(x,t)$ at two arbitrary points $x$ and $x'$ along the length of the quasicondensates, and is experimentally measurable \cite{langen2015b}. In the quasicondensate regime considered throughout this work, wherein the density fluctuations are suppressed and can be neglected, the two-point correlation function $C(x,x',t)$ simplifies to 
\begin{equation}\label{eq:PCF}
C(x,x',t) \approx \langle \exp[i\Delta \phi(x,t)-i\Delta \phi(x',t)] \rangle,
\end{equation} 
where $\Delta \phi(x,t) = \phi_{1}(x,t) - \phi_{2}(x,t)$ is the relative phase between the two quasicondensates. As such, it quantifies the degree of correlation between the local relative phases of the two quasicondensates at two arbitrary points $x$ and $x'$ \cite{whitlock2003,betz2011,langen2013b}. For all example systems treated in this work, both Eq.~\eqref{eq: C2} and Eq.~\eqref{eq:PCF} gave quantitatively very similar results, confirming that the density fluctuations in our systems are indeed negligible and hence implying that the presence of any correlation in Eq.~\eqref{eq: C2} can be attributed purely to phase correlation or phase locking due to the tunnel coupling.

In Fig.~\ref{fig: DiffNT-C2}, we plot the {initial equilibrium ($t=0_-$) and final relaxed ($t=t_f$) correlation functions} $C(x,x'=0,t)$, for {initially fully balanced quasicondensates} (i.e., the same initial parameters as in Fig.~\ref{fig: momdens}) and for two different values of the tunnel-coupling strength $J$: one corresponding to a strong {cross-coupling} regime with $\tilde{J}=3\tilde{\omega}$ satisfying $l^{(J)} \ll l_{0,j}^{(\phi)}$ (dotted yellow line) and the other to an intermediate {cross-coupling} regime with $\tilde{J}=\tilde{\omega}/4$ (dashed red line). 
{We recall that the lengthscale $l^{(J)} = \sqrt{\hbar/(4mJ)}$, associated with the coupling strength $J$, represents the characteristic distance over which the tunnel coupling restores a spatially constant relative phase $\Delta\phi(x,t) = \phi_{1}(x,t) - \phi_{2}(x,t)$.}
As we see from the figure, the tunnel coupling establishes stronger correlation between the relative phases of the two relaxed quasicondensates at time $t_f=200/\omega$, which can be referred to as phase locking, {$\Delta \phi=0$} \cite{betz2011,schweigler2017}. The correlation has larger amplitude  on a longer range for  stronger tunnel-coupling as expected. {The correlations eventually decay in all cases as soon as the relative distance approaches the size of the cloud, which is well approximated by the Thomas-Fermi radius and which is equal to $R_{1,2}^{(TF)}\simeq 14.8l_{\mathrm{ho}}$ in this example.}

Returning to the phase coherence length within the relaxed quasicondensates, we point out that an increasingly stronger tunnel-coupling does not necessarily lead to an increasingly longer phase coherence length within each quasicondensate. This is because of the simultaneous increase in the temperature of the relaxed coupled system: higher temperatures would generally reduce the phase coherence length, however, in the coupled setup such a reduction competes with the increase of phase coherence length due to phase locking so that the overall effect will depend on system parameters.

As a final note, in the inset of Fig.~\ref{fig: DiffNT-C2} we plot the relative phase correlation function $C(x,x'=0,t)$ as a function of time, monitoring it at six different fixed positions $x$ from the trap center, starting from $x = 2.4l_{\mathrm{ho}}$ (topmost curve) to $x = 14.4l_{\mathrm{ho}}$ (bottommost curve). {All these positions (relative distances) are within the bulk of the quasicondensates, with $x$ being smaller than the Thomas-Fermi radius $R_{1,2}^{(TF)}\simeq14.8l_{\mathbf{ho}}$. As we see, the correlation functions $C(x,x'=0,t)$ for the shorter relative distances, where the phase locking is stronger, reach their equilibrium values by approximately $t \lesssim 7/\omega$. For larger relative distances, the correlations appear to reach their thermal equilibrium values somewhat earlier, but the correlation strength (or the strength of phase locking) here is much weaker regardless. This short relaxation time in the bulk of the quasicondensates, where the low energy particles reside, further confirms our earlier observation that any residual dynamics seen in $E_{\mathrm{imb}}(t)$ and $N_{\mathrm{imb}}(t)$ past this time (as in Fig.~\ref{fig: DiffNT}) is due to the high-energy particles that occupy the tails (rather than the bulk) of the quasicondensates, where phase locking is essentially absent.}

\section{Summary}
\label{summary}

In conclusion, we have studied the relaxation dynamics of two initially independent (uncoupled) 1D quasicondensates following a sudden quench of the strength of tunnel coupling between the quasicondensates from zero to a constant finite final value. We observe that the coupled quasicondensates relax to a final higher-temperature equilibrium state described by a thermal Gibbs ensemble. 
 
{The dynamics of the particle number imbalance between the quasicondensates is characterized by the flow of low energy particles from the colder to the hotter quasicondensate and the flow of high energy particles in the opposite direction.  If the number of particles in these opposite flows is equal to each other, there will be net energy flow from the hot quasicondensate to the cold one, as expected. This situation was illustrated in the example of Fig.~\ref{fig: DiffNT}\,(a). If, however, the number of low energy particles flowing from the colder to the hotter quasicondensate exceeds the number of high energy particles flowing in the opposite direction, one can have a balanced situation with no net energy flow between the quasicondensates even though the initial temperatures were different. This scenario was illustrated in Fig.~\ref{fig: DiffNT}\,(b). Finally, as in the example of Fig.~\ref{fig: DiffNT}\,(c), if the number of low energy particles flowing from the colder to the hotter quasicondensate significantly exceeds the number of high energy particles flowing in the opposite direction, we observed net energy flow from the colder quasicondensate to the hotter one. In all instances, where the energy flows in the seemingly counterintuitive direction (i.e., from the colder system to the hotter one), it is identified as chemical work (rather than heat) governed by the difference in the initial chemical potentials.}  
{Similar seemingly counterintuitive flows have been observed by the ETH Zurich group, who have investigated  transport between two reservoirs of a degenerate Fermi gas with different chemical potentials and temperatures coupled by a channel~\cite{JPB2013,DH2017,PhysRevX.11.021034}.}

We also analysed the phase coherence properties of the coupled quasicondensates and found that, even though the final temperature of the system is higher (which for the same independent quasicondensates would imply a shorter thermal phase coherence length within each quasicondensate), the coherence length of the relaxed coupled quasicondensates can become larger than the initial coherence length. This is explained by the fact that the tunnel coupling acts a phase locking mechanism and is equivalent to an additional degree of freedom in the system, which generally favours establishment of phase coherence over a longer range. The effect is, however, in competition with the trend of reduction of phase coherence length with temperature.

{It would also be interesting to incorporate a tunnel-coupled system like the one studied here into a full thermodynamic cycle of a prototype quantum gas heat engine and simulate its performance. In such a setup, it would be preferable to consider large particle number imbalance between the two quasicondensates so that the smaller quasicondensate can be regarded as the working fluid, whereas the larger quasicondensate as the heat bath with which the working fluid equilibrates after the work strokes in, e.g., an Otto engine cycle.}

\begin{acknowledgments}
The authors acknowledge support by the Australian Research 
 Council Discovery Project Grants No. DP170101423 (K.V.K), No. DP190101515 (K.V.K.), and the Australian Research Council Centre of Excellence for Engineered Quantum Systems (EQUS, CE170100009) (M.J.D.). 
\end{acknowledgments}

\appendix 

\section{Dimensionless form of the SPGPE}
\label{dimensionless}

In order to arrive at the dimensionless form of the SPGPE, which will then depend explicitly on the earlier introduced dimensionless parameter $\chi_{0,j}$, {Eq. \eqref{eq: chi0}}, for the $j$th quasicondensate (see Ref.~\cite{Thomas-2021} for further details), we introduce the dimensionless coordinate $\xi = x/x_{0}$, time $\tau = t/t_{0}$, and field $\varphi_{j}(\xi,\tau) = \psi_{j}^{(\mathcal{C})}(x,t)/\psi_{0}$ with the corresponding length-, time-, and field-scales in terms of the temperature of the first quasicondensate:
\begin{align*}
x_{0} =& \biggl( \frac{\hbar^{4}}{m^{2}gk_{\mathrm{B}}T_{1}} \biggr)^{1/3}, \\
t_{0} =& \frac{mx_{0}^{2}}{\hbar} = \biggl( \frac{\hbar^{5}}{mg^{2}k_{\mathrm{B}}^{2}T_{1}^{2}} \biggr)^{1/3},\\
\psi_{0} =& \biggl( \frac{mk_{\mathrm{B}}^{2}T_{1}^{2}}{\hbar^{2}g} \biggr)^{1/6}{=\bigg( \frac{\hbar}{t_0g}\bigg)^{1/2}}.
\end{align*}

With these scaling factors, and the role of the energy scale taken by {$E_0=\hbar/t_0$}, the SPGPE {[Eq.~\eqref{SPGPE1} of the main text]} acquires the following compact form:
\begin{equation}\label{eq: SPGPE}
d\varphi_{j} = \mathcal{P}^{(\mathcal{C})} \bigl\{ \bigl[ -i\mathcal{L}_{j} + \tilde{\kappa}_{\mathrm{th}} \bigl( \tilde{\mu}_{j} - \mathcal{L}_{j} \bigr) \bigr] \varphi_{j}d\tau + d\tilde{W}_{j} \bigr\},
\end{equation}
where the {dimensionless nonlinear operator $\mathcal{L}_{j}$ is obtained from Eq.~\eqref{GP-operator} via $\mathcal{L}_{j}=\mathcal{L}^{(C)}_{j}/E_0$ and is given by}
\begin{equation}\label{GP-operator2}
\mathcal{L}_{j}=  -\frac{1}{2}\frac{\partial^{2}}{\partial \xi^{2}} + \frac{1}{2}\tilde{\omega}^{2}\xi^{2} + |\varphi_{j}|^{2}.
\end{equation}
Here, $\tilde{\omega} = \omega t_{0}$ is the dimensionless trap frequency, whereas $\tilde{\mu}_{j} = \mu_{j}/E_0$ is the dimensionless chemical potential of the $j$-th quasicondensate. In addition, $\tilde{\kappa}_{\mathrm{th}} = \hbar\Gamma_1/k_{\mathrm{B}}T_1$ is the rescaled growth rate, {where we have} used the freedom of choice of the numerical value of the growth rate $\Gamma_2$ to relate it to $\Gamma_1$ {via $\Gamma_2=\Gamma_1\alpha$, where $\alpha=T_2/T_1$ is the ratio of temperatures of the two quasicondensates}. The terms {$d\tilde{W}_{1}=dW_1/\psi_0$} and $d\tilde{W}_{2} = \sqrt{\alpha} d\tilde{W}_{1}$ are complex white noises satisfying $\langle d\tilde{W}_{1}^{*}(\xi,\tau) d\tilde{W}_{1}(\xi',\tau) \rangle = 2\tilde{\kappa}_{\mathrm{th}}\delta(\xi-\xi')d\tau$, {from Eq.~\eqref{noise-corr}}. 

The dimensionless form of the coupled GPEs \eqref{GPE}, on the other hand, reads as
\begin{eqnarray}
\frac{\partial \varphi_{1}(\xi,\tau)}{\partial \tau}&=& - i\mathcal{L}_1\varphi_{1}(\xi,\tau) + i \tilde{J} \varphi_{2}(\xi,\tau), \nonumber \\
\frac{\partial \varphi_{2}(\xi,\tau)}{\partial \tau}&=& - i\mathcal{L}_2\varphi_{2}(\xi,\tau) + i \tilde{J} \varphi_{1}(\xi,\tau),
\label{GPE2}
\end{eqnarray}
$\tilde{J} = \hbar J/E_0=Jt_0$ is the dimensionless tunnel-coupling. In the above dimensionless form of the SPGPE and GPE, the nonlinearity constant {in front of the $|\varphi_j|^2$ term in Eq.~\eqref{GP-operator2} is always equal to unity (unlike the respective term in Eq.~\eqref{GP-operator}, proportional to the interaction strength $g$)}, and the normalization condition that gives the total number of particles in the $j$th quasicondensate reads {$N_j\!=\!\int \langle |\psi_j^{(\mathcal{C})}(x,t)|^2 \rangle dx\!=\!\psi_0^2x_0 \tilde{N}_j$, where $\tilde{N}_j\!=\!\int \langle |\varphi_j(\xi,\tau)|^2 \rangle d \xi$.}

In the Thomas-Fermi (TF) limit of an inverted parabolic density profile, the chemical potential of a harmonically trapped quasicondensate is given by $\mu_j=g\rho_{0,j}$, and thus {the dimensionless chemical potential $\tilde{\mu}_1$ can be expressed in terms of the dimensionless parameter $\chi_{0,1}$, {Eq. \eqref{eq: chi0}}, as
\begin{equation}
\tilde{\mu}_1=1/\chi_{0,1}^{2/3}, 
\label{eqn:mu_bar}
\end{equation} 

 According to this relation, the dimensionless chemical potential $\tilde{\mu}_1$ can be interchanged with $1/\chi_{0,1}^{2/3}$. This implies then that the SPGPE for a single quasicondensate depends on a  nontrivial combination of both the dimensionless interaction strength ($\gamma_{0,j}$) and temperature ($\mathcal{T}_{j}$), rather than on two independent parameters. }
{Given that the full density profile in the TF approximation is given by $\rho_1(x)=(\mu_1-\frac{1}{2}m\omega^2x^2)/g$, for $|x|<R_1^{(TF)}$ (and $\rho_1(x)=0$ otherwise), where $R_{1}^{(TF)}=\sqrt{2\mu_1/m\omega^2}$ is the TF radius and $\mu_1=g\rho_{0,1}$ is the global chemical potential, the peak density $\rho_{0,1}$ can be evaluated from the normalization condition $N_1=\int_{-R_{1}^{(TF)}}^{R_{1}^{(TF)}}\rho_1(x)\,dx$. This gives $\rho_{0,1}\!=\!(9m\omega^2N_1^2/32g)^{1/3}$, and therefore the dimensionless parameter $\chi_{0,1}$ from Eq.~\eqref{eq: chi0}} can be explicitly rewritten as $\chi_{0,1}\!=\!4\sqrt{2}k_BT_1/(3N_1\hbar\omega)$ [using $N_1\!=\!\psi_0^2x_0 \tilde{N}_1$ and $\tilde{N}_1\!=\!4 \sqrt{2}\tilde{\mu}_1^{3/2}/(3\tilde{\omega})\!=\!4 \sqrt{2} /(3\chi_{0,1}\tilde{\omega})$], and similarly for $\chi_{0,2}\!=\!4\sqrt{2}k_BT_2/(3N_2\hbar\omega)$. Beyond the TF limit, the dimensionless chemical potential $\tilde{\mu}_1$ can still be interchanged with $\chi_{0,1}$ as an input parameter, with the understanding that the simple relationship $\mu_1=g\rho_{0,1}$ between $\mu_1$ and the TF peak density $\rho_{0,1}$ is now only approximate, whereas the exact relationship has to be determined numerically \emph{a posteriori}, {using $\rho_1(x) =\langle |\psi_{1}^{(\mathcal{C})}(x,t)|^2\rangle$. We note, however, that in all our numerical examples, the equilibrium peak densities were always very close to the TF peak density of $\rho_{0,1}=\mu_1/g$.}

{Recalling that for two independent condensates, we had to introduce two additional dimensionless parameters for the ratios of the initial temperatures and chemical potentials, $\alpha=T_2/T_1$ and $\beta=\mu_2/\mu_1$, we thus conclude that the initial state of our (uncoupled) system can be completely characterized by just four dimensionless parameters, $\chi_{0,1}$ (or $\tilde{\mu}_1\!=\!1/\chi_{0,1}^{2/3}$), $\tilde{\omega}$, $\alpha$, and $\beta$. In doing so, we further note that as $\mu_j= g\rho_{0,j}$ in the Thomas-Fermi limit, one has $\chi_{0,2} = \bigl( \alpha/\beta^{3/2} \bigr)\chi_{0,1}$ (with $\tilde{\mu}_2=\beta\tilde{\mu_1}$), and therefore the value of $\chi_{0,2}$ is determined by $\chi_{0,1}$, $\alpha$, and $\beta$. 
The post-quench dynamics of the coupled quasicondensates, on the other hand, requires a fifth parameter---the dimensionless tunnel coupling $\tilde{J}$. }


%

\end{document}